\documentclass[peerreview,onecolumn,11pt,draftclsnofoot]{IEEEtran}
\usepackage{amsmath}
\usepackage{amsfonts}
\usepackage{epsfig}
\usepackage{amssymb}
\usepackage{graphicx}
\usepackage{amssymb,amsmath}
\usepackage{cite}
\usepackage{color,soul}

\begin{document}
\title{Worst-Case Robust Distributed Power Allocation in Shared Unlicensed Spectrum}

\author{Saeedeh Parsaeefard and
        Ahmad R. Sharafat
\thanks{Manuscript received June 23, 2011. This work was supported in part by Tarbiat Modares University, and in part by Iran Telecommunications Research Center, Tehran, Iran, under PhD Research Grant TMU 88-11-124.

The authors are with the Department of Electrical and Computer
Engineering, Tarbiat Modares University, P. O. Box 14155-4838,
Tehran, Iran. Corresponding author is A.~R.~Sharafat (e-mail:
sharafat@modares.ac.ir).}}

\maketitle

\begin{abstract}
This paper considers non-cooperative and fully-distributed power-allocation for selfish transmitter-receiver pairs in shared unlicensed spectrum when normalized-interference to each receiver is uncertain. We model each uncertain parameter by the sum of its nominal (estimated) value and a bounded additive error in a convex set, and show that the allocated power always converges to its equilibrium, called robust Nash equilibrium (RNE). In the case of a bounded and symmetric uncertainty region, we show that the power allocation problem for each user is simplified, and can be solved in a distributed manner. We derive the conditions for RNE's uniqueness and for convergence of the distributed algorithm; and show that the total throughput (social utility) is less than that at NE when RNE is unique. We also show that for multiple RNEs, the social utility may be higher at a RNE as compared to that at the corresponding NE, and demonstrate that this is caused by users' orthogonal utilization of bandwidth at RNE. Simulations confirm our analysis.

\end{abstract}

\section{Introduction}

\IEEEPARstart{O}{pportunistic} spectrum sharing where multiple wireless transmitters and receivers share the same portion of bandwidth is a promising approach for improving spectrum efficiency in future wireless communication systems \cite{spectrumsharing}. In this setup, each user aims to maximize its utility subject to its power limit and other regulatory restrictions, such as spectrum mask and/or the maximum amount of interference \cite{Haykin}.

Because of the inherently decentralized and competitive nature of allocating power to users, game theory \cite{NEexistence} is an appropriate tool for analyzing such systems \cite{MIMOcognitivescutari,Robusthaykin,ScutariVI,ProbabilisticIWFA,Robustnew}. In this context, the power allocation problem is formulated as a strategic non-cooperative game, where each user is considered as a rational player that competes with others by choosing a transmission strategy for maximizing its own utility, defined as its throughput. In a game, Nash equilibrium (NE) is a state, consisting of the strategy space and utility values, at which there is no incentive for any player to change its strategy, provided that other players' strategies are not changed. As such, one needs to derive the conditions for existence and uniqueness of NE, to develop a distributed algorithm for reaching NE, and to examine the convergence conditions for that algorithm.

The well known iterative water-filling algorithm (IWFA) for reaching NE of a power allocation game, as well as the conditions for existence of NE and convergence conditions of IWFA are proposed in \cite{Yu}. Besides, in \cite{MIMOcognitivescutari, ScutariVI, scutari5,25, Luo2, Nash1, Nash2,SimultanousWFA,spectrumsharing}, other distributed water-filling based algorithms and conditions for their convergence are studied, and the sufficient conditions for existence and uniqueness of NE under different power constraints and system models are derived. In many of the existing power allocation schemes (e.g., in \cite{25, Luo2, Nash1, Nash2,SimultanousWFA,VIintroduction,Unifiedscutari,PCbookchiang}), it is assumed that the channel state information, and interference from other users plus noise (IN) are available to each user. However, this assumption may not be valid in practice, due to the time-varying environment and inaccurate measurements, resulting in uncertainties in parameter values.

Motivated by the aforementioned challenge, we wish to develop a robust game-theoretic approach for tackling such
uncertainties in the power allocation problem. Generally, to model the uncertainty, each uncertain parameter is assumed to be the sum of its nominal (or estimated value) and an additive error \cite{Gershman}. Based on the available information on the additive error, a robust approach is considered. In the literature, there are two major approaches for introducing robustness \cite{selecectedrobust,Gershman}: The Bayesian approach where the statistics of errors are considered and the network's performance is probabilistically guaranteed; and the worst-case approach where a deterministic value bounded with a given probability in a closed region (called the uncertainty region), is considered for the additive error, and the network's performance is guaranteed for any realization of uncertainty within the uncertainty region. In this paper, for brevity, we use the term ``bounded'' instead of the term ``bounded with a given probability''. Both of the above mentioned approaches have been applied to the power allocation problem in spectrum sharing environments. The term ``worst-case approach'' is used in the literature to denote the fact that for any realization of uncertainty in the bounded region, including the bounds (i.e., the worst-case for the uncertainty), the network's performance is guaranteed by the proposed robust approach.

In \cite{ProbabilisticIWFA}, a probabilistic robust IWFA is proposed, where IN levels are uncertain; and by assuming a uniform probability density function (pdf) for uncertainty in IN levels, the power allocation problem is converted to the conventional IWFA, but the IN level is multiplied by a factor that corresponds to the stochastic nature of uncertainty. In \cite{Robustnew}, the worst-case robust optimization theory is used when uncertainty in the channel state information between users is bounded in an ellipsoid region to derive the conditions for NE's existence and uniqueness, and a distributed algorithm that needs additional message passing between users is proposed. Compared to pervious works, our main contributions in this paper are two folds. First, we propose a simple robust distributed power allocation scheme that does not need any additional message passing in the system; and second, we analyze and compare the equilibrium of the robust approach with that of the power allocation problem with no uncertainties.

In doing so, we assume that the IN level for each user normalized by its direct channel gain is uncertain, and as in \cite{Robusthaykin,Robustnew}, model each uncertain parameter by the sum of its nominal (estimated) value plus a bounded additive error, the collection of which for all instances of error form the uncertainty region. The corresponding game is based on the robust optimization theory \cite{Robustgame}, where each user obtains its transmit power level that maximizes its throughput in the worst-case instance in the uncertainty region \cite{selecectedrobust}, and the equilibrium of such a game is called the robust Nash equilibrium (RNE). We will show that when the uncertainty region is a closed convex set, RNE always exists, but a closed form solution for RNE may not be obtainable for some forms of the uncertainty region. We focus on bounded symmetric uncertainties, and utilize the framework in \cite{Nash1,Nash2} to obtain the necessary conditions for uniqueness of RNE. Moreover, to reach RNE, we will apply the totally asynchronous and distributed power allocation algorithm introduced in \cite{scutariasynch}, and obtain the condition for its convergence.

Another important subject in this paper is to obtain the difference between RNE of the robust game with uncertain parameter values and NE of the nominal game (the game with complete information). We show how uncertainty affects the total throughout, i.e., the social utility of users at RNE as compared to that of nominal game. We also show that when RNE is unique, the strategy of users as well as the social utility of the game are decreasing functions of the bound on the uncertainty region. Besides, we demonstrate that uncertainty reduces the total throughput of users when RNE is unique, but not necessarily for multiple RNEs. When multiuser interference is high, multiple RNEs may exist in IWFA-based algorithms for the robust game. In such cases, uncertainty may lead to a more orthogonal power allocation at one RNE, resulting in less interference between users and consequently, higher total throughput of users as compared to the total throughput of the nominal game.

This paper is organized as follows. In Section II, we present the system model of a spectrum sharing environment, and formulate a robust game for the power allocation problem, when users' channel state information and IN levels are uncertain. In Section III, we investigate the existence and uniqueness conditions of RNE in the proposed game, and study the effect of uncertainties on the total throughput at RNE as compared to the same at NE of the complete information game. Our distributed algorithm is proposed in Section IV, followed by simulation results in Section V, and conclusions in Section VI.

\section{System Model and Problem Statement}

We consider a set of $\mathcal{M}=\{1,\cdots, M\}$ transmitter and receiver pairs sharing $\mathcal{K}=\{1,\cdots,K\}$ orthogonal narrow band sub-channels. The bandwidth of each sub-channel is much less than the coherence bandwidth of the wireless channel, meaning that the channel response of each sub-channel is flat. The transmit power vector of the $i^{\scriptsize{\textnormal{th}}}$ user over all sub-channels is $\mathbf{p}_{_i}=[p^{1}_{i},\cdots,p^{K}_{i}]$, where $p^{k}_{i}$ is the transmit power of the $i^{\scriptsize{\textnormal{th}}}$ user in the $k^{\scriptsize{\textnormal{th}}}$ sub-channel. The received signal power at the corresponding receiver in sub-channel $k$ is
\begin{equation}\label{recive}
    r^{k}_{i}=\sum^{M}_{j=1} p^{k}_{j}h^{k}_{ji}+\sigma_{i}^{k}, \quad
    \forall j \in \mathcal{M} , \quad  k \in \mathcal{K},
\end{equation}
where $h^{k}_{ji}$ is the fading sub-channel gain from the $j^{\scriptsize{\textnormal{th}}}$ transmitter to the
$i^{\scriptsize{\textnormal{th}}}$ receiver on the
$k^{\scriptsize{\textnormal{th}}}$ sub-channel, and
$\sigma_{i}^{k}$ is the noise power in the $k^{\scriptsize{\textnormal{th}}}$ sub-channel of the
$i^{\scriptsize{\textnormal{th}}}$ user. At receiver $i$, signal-to-interference-plus-noise-ratio (SINR) in sub-channel $k$ is
      $\gamma^{k}_{i}= \frac{p^{k}_{i}}{s^{k}_{i}}$,
where $s^{k}_{i}=\frac{\sum_{j\neq i}
p^{k}_{j}h^{k}_{ji}+\sigma_{i}^{k}}{h_{ii}^{k}}$ denotes the induced interference plus noise to user $i$ normalized by its direct channel gain. We assume that at the receiver of each user, the value of IN in each sub-channel and the corresponding direct channel gain $h_{ii}^k$ are measured. The receiver calculates the value of $s_{i}^{k}$ which is the normalized interference caused due to the direct channel gain, and sends it to the corresponding transmitter via the feedback channel. Each transmitter uses the received value of $s_{i}^{k}$ to calculate the power level in each sub-channel. At the receiver, the value of SINR in each sub-channel can be calculated by $\frac{p^{k}_{i}}{s^{k}_{i}}$, and the achievable rate for the $i^{{\scriptsize{\textnormal{th}}}}$ receiver is obtained by $R_i= \sum_{k=1}^{K}\log(1+ \gamma^{k}_{i})$. The transmit power of each user is subject to the following constraints:
\begin{description}
    \item [$C_1$:] The total transmit power of the $i^{\scriptsize{\textnormal{th}}}$ user over all sub-channels is limited by its maximum power budget, i.e.,
    $\sum_{k=1}^{K} p^{k}_{i}\leq
    p_{i}^{\scriptsize{\textnormal{max}}}$.
    \item [$C_2$:]The transmit power of each user on each sub-channel is limited, i.e.,
$0\leq p^{k}_{i}\leq p^{k}_{\scriptsize{\textnormal{mask}}}$, where $p^{k}_{\scriptsize{\textnormal{mask}}}$ is the spectral mask on sub-channel $k$.
\end{description}
We assume that users cannot perform interference cancelation \cite{Nash1}.

\subsection{Game Formulation}

In noncooperative spectrum sharing in unlicensed bands, each user aims to maximize its own utility, defined as its throughput, subject to $C_1$ and $C_2$. In contrast to cognitive radio networks where primary (licensed) users have absolute priority over secondary (unlicensed) users in utilizing the frequency spectrum, in noncooperative spectrum sharing in unlicensed bands \cite{spectrumsharing}, all users are treated the same, i.e., there is no priority among users. In such cases, noncooperative strategic game theory \cite{NEexistence} is an appropriate tool for analyzing such a greedy behavior by each users where all users act independently to achieve their objectives without any coordination. The strategic game consists of three elements: the set of players, the utility of each player, and the action set of players, where the action of each player is its strategy for optimizing its utility.

For our power allocation problem, the set of players is the set of all transmitters and receivers in the wireless channel denoted by $\mathcal{M}$, the action of each player is its transmit power over $K$ sub-channels denoted by $\mathbf{p}_i$, and the utility of each user is its achieved throughput over all $\mathcal{K}$ sub-channels. This game is denoted by $\mathcal{G} \triangleq \langle \mathcal{M},\{ \mathcal{P}_{i}\}_{i \in \mathcal{M}}, \{u_{i}\}_{i\in \mathcal{M}}\rangle $, where $\mathcal{P}_{i}$  is the action set of player $i$ defined as
\begin{equation}\label{Powerstratgy}
    \mathcal{P}_{i} \triangleq \{ \mathbf{p}_{_i} |  \sum_{k=1}^{K} p^{k}_{i} \leq  p_{i}^{\scriptsize{\textnormal{max}}}, \quad 0\leq p^{k}_{i}\leq  p^{k}_{\scriptsize{\textnormal{mask}}}, \quad  \forall k \in \mathcal{K}\},
\end{equation}
and $u_{i}(\mathbf{p})$ is the utility of user $i$, and depends on the chosen strategy vector of all users, i.e.,
\begin{equation}\label{utilitydefnitionn}
   u_{i}(\mathbf{p})=\sum_{k=1}^{K}\log(1+\frac{p^{k}_{i}}{s^{k}_{i}}),
\end{equation}
where $\mathbf{p}=[\mathbf{p}_{_i},\mathbf{p}_{_{-i}}]$, and $\mathbf{p}_{_{-i}}=[\mathbf{p}_{_1},\cdots,\mathbf{p}_{_{i-1}},\mathbf{p}_{_{i+1}},\cdots,\mathbf{p}_{_M}]$ is a vector of the actions of all users except user $i$.

In the noncooperative strategic game, each player maximizes its own utility based on actions of other players. Since the utility function of each user for the power allocation problem is achieved throughput i.e. (\ref{utilitydefnitionn}), and its strategy space is defined by (\ref{Powerstratgy}), optimization of the allocated power is stated by the following problem
\begin{eqnarray}\label{IWFutility1}
  \max_{\mathbf{p}_{_i}, \mathbf{p}_{_{-i}}} & \sum_{k=1}^{K}\log(1+\frac{p^{k}_{i}}{s^{k}_{i}})
 \end{eqnarray}
\[\mathrm{subject~to} \left\{
\begin{array}{l l}
 C_1: \sum_{k=1}^{K} p^{k}_{i} \leq  p_{i}^{\scriptsize{\textnormal{max}}} \nonumber \\
C_2: 0\leq p^{k}_{i}\leq  p^{k}_{\scriptsize{\textnormal{mask}}} \nonumber \\
\end{array} \right. \]
The commonly used concept for analyzing the outcome of this type of game-theoretic schemes is the Nash equilibrium (NE). NE represents a strategy set of all players in the game where no player can improve its utility by unilaterally changing its action \cite{NEexistence}. In the power allocation problem, the strategy profile $\mathbf{p}^{*}=\{\mathbf{p}^{*}_{_1},\cdots,\mathbf{p}^{*}_{_M}\}$ is NE for the game $\mathcal{G}$ if
\begin{equation}\label{NEpoint}
 u_{i}(\mathbf{p}^{*}_{_i},\mathbf{p}^{*}_{_{-i}})\geq
 u_{i}(\mathbf{p}_{_i},\mathbf{p}^{*}_{_{-i}}), \quad \forall \mathbf{p}_{_{i}} \in \mathcal{P}_i, \quad \forall i \in \mathcal{M}.
\end{equation}
One approach for reaching NE is to use the best response dynamic algorithm \cite{Nash1,NEexistence,mihaelastructure}. For the power allocation problem, the best response of user $i$, given the transmit power levels of other users, is the optimal solution to (\ref{IWFutility1}) \cite{Nash1}. Since (\ref{IWFutility1}) is a convex optimization problem with respect to $\mathbf{p}_i$, its solution can be obtained by the Lagrange dual function. The dual function of (\ref{IWFutility1}) is
\begin{equation}\label{Lagranjain}
L_i(\mathbf{p}_i, \lambda_i)=\sum_{k=1}^{K}\log(1+\frac{p^{k}_{i}}{s^{k}_{i}}) +\lambda_i \times (\sum_{k=1}^{K} p^{k}_{i} - p_{i}^{\scriptsize{\textnormal{max}}}),
\end{equation}
where $\lambda_i>0$ is the dual variable for user $i$ that satisfies its $C_1$, i.e.,
\begin{eqnarray}\label{optimaloptlambda}
\lambda_{i} \times (\sum^{K}_{k=1}p_{i}^{k}-
p_{i}^{\scriptsize{\textnormal{max}}})=0,  \qquad\forall i \in
\mathcal{M}.
\end{eqnarray}
The allocated power to each user can be obtained by the first order optimality condition of the Lagrange dual function, i.e., via $\frac{\partial L_i(\mathbf{p}_i, \lambda_i)}{\partial \mathbf{p}_i}=0$ as
\begin{equation}\label{IWFA}
   p_{i}^{k}=[\frac{1}{\lambda_{i}}-s^{k}_{i}]^{p^{k}_{\scriptsize{\textnormal{mask}}}}_{0},
\end{equation}
where $[x]^{b}_{a}$  for $a<b$ denotes the Euclidean projection
of $x$ onto the interval $[a,b]$, i.e., $[x]^{b}_{a}=a$ if $x<a$, $[x]^{b}_{a}=ax$ if $a<x<b$, and $[x]^{b}_{a}=b$ if $b<x$. This best response solution to (\ref{IWFutility1}) for each player is called the water-filling solution \cite{boydconvexbook, Nash1}, and the iterative algorithm utilized by each player to reach the equilibrium point is called the iterative water filling algorithm (IWFA) \cite{Yu,Nash1, Nash2}.

To analyze the performance of the game, one needs to investigate the existence and uniqueness of NE, propose a distributed algorithm to reach NE, and obtain its convergence conditions. For the power allocation problem, since the utility function $u_{i}(\mathbf{p})$ is concave on $\mathbf{p}_{_i}$, and the action set of each user is closed and convex, the game $\mathcal{G}$ has a nonempty solution set for any set of channels, spectral mask constraints, and transmit power levels. Hence, a NE always exists \cite{NEexistence}. However, uniqueness of NE for the power control game cannot be guaranteed in the general case. To analyze the uniqueness of NE in the power allocation problem, different conditions have been derived for the game $\mathcal{G}$ in the literature, depending on the interference channel gains and noise levels in the system as in \cite{Yu,Nash1,mihaelastructure,24,25} via best-response algorithms and principles of contraction mapping. Briefly, in all existing literature (e.g., in \cite{Yu,Nash1,Unifiedscutari}), it is shown that if multiuser interference is low, the game has a unique NE, and when multiuser interference is high, the game has multiple NEs. In this paper, to analyze our robust game, we utilize the framework introduced in \cite{Nash1} to derive the condition for uniqueness of RNE.

\subsection{Robust Counterpart Game Formulation}

In the above formulation for game $\mathcal{G}$, it is assumed that the exact value of $s_{i}^{k}$ for each sub-channel is available to the respective receiver with no error. This value is sent to the corresponding user via the feedback channel. However, due to the dynamic nature of spectrum sharing environments manifested in channel variations, users' movements, entering new users in the system, as well as the delay in the feedback channel, errors are introduced in $s_{i}^{k}$, which invalidate the assumption that the error-free value of $s_{i}^{k}$ is available to the the respective user. This means that power allocation based on (\ref{IWFA}) cannot guarantee the expected optimal utility of each user in reality.

To tackle this problem, in our formulation, we assume that each uncertain parameter $s_i^k$ is modeled by the sum of its estimate (nominal value) and an uncertain term, i.e.,
$\tilde{s}^{k}_{i} = \bar{s}^{k}_{i}+\hat{s}^{k}_{i}$,
where $\tilde{s}^{k}_{i}$, $\bar{s}^{k}_{i}$ and $\hat{s}^{k}_{i}$ are the actual, nominal
value and error in  $s^{k}_{i}$ for the $i^{\text{th}}$
user, respectively. Similar to the robust optimization in game theory \cite{Robustgame}, we assume that the statistics of uncertain parameters are unknown, but the distances between the nominal and the actual values of uncertain parameters are bounded with a given probability, i.e., 
 \begin{equation} \label{uncertaintyregion1}
 \mathcal{R}_{s} = \{ \tilde{\mathbf{s}}_{i}  \in  \mathcal{R}_{s} |  \| \tilde{\mathbf{s}}_{i}-\bar{\mathbf{s}}_{i} \| \leq \varepsilon_i\}  , \quad
  \forall  i \in \mathcal{M},\quad \forall  k \in \mathcal{K}.
 \end{equation}
where $\|\cdot\|$  is any definition of norm for the vector space \cite{boydconvexbook}, $\tilde{\mathbf{s}}_{i}=[\tilde{s}_{i}^{1},\cdots,\tilde{s}_{i}^{K}]$, $\hat{\mathbf{s}}_{i}=[\hat{s}_{i}^{1},\cdots,\hat{s}_{i}^{K}]$, $\bar{\mathbf{s}}_{i}=[\bar{s}_{i}^{1},\cdots,\bar{s}_{i}^{K}]$, and $\varepsilon_i$ is the bound on the distance between the nominal and the actual value, and $\varepsilon_i$ is the size of the uncertainty region. Large values of $\varepsilon_{i}$ expands the uncertainty region, meaning higher uncertainties. Note that this approach can also be applied when the uncertain parameters behave stochastically, and the bound on uncertainty can be obtained from the pdf of the error (see Appendix A).

The uncertain parameter is a new optimization variable in the optimization problem of each user, which modifies the utility function of each user to
\begin{equation}\label{utilitydefnition}
   \widetilde{u}_{i}(\mathbf{p}, \tilde{\mathbf{s}}_i)=\sum_{k=1}^{K}\log(1+\frac{p^{k}_{i}}{\tilde{s}^{k}_{i}}),
\end{equation}
and the robust counterpart of (\ref{IWFutility1}) is changed \cite{selecectedrobust} to
\begin{equation}\label{IWFrobustcounterpart}
  \max_{\mathbf{p}_{i}, \mathbf{p}_{-i}} \min_{\tilde{\mathbf{s}}_{i}\in \mathcal{R}_{s} }\sum_{k=1}^{k}\log(1+\frac{p^{k}_{i}}{\tilde{s}^{k}_{i}})
 \end{equation}
 \[\mathrm{subject~to} \left\{
\begin{array}{l l}
 C_1: \sum_{k=1}^{K} p^{k}_{i} \leq  p_{i}^{\scriptsize{\textnormal{max}}} \nonumber \\
 C_2: 0 \leq p^{k}_{i}\leq  p^{k}_{\scriptsize{\textnormal{mask}}} \nonumber \\
\end{array} \right. \]
We define the corresponding robust game for the objective function (\ref{IWFrobustcounterpart}) by $\widetilde{\mathcal{G}}
\triangleq \langle \mathcal{M},\{\mathcal{P}_{i}\}_{i \in \mathcal{M}}, \{\widetilde{u}_{i}\}_{i\in\mathcal{M}} \rangle$.
For both games $\mathcal{G}$ and $\widetilde{\mathcal{G}}$, the sets $\mathcal{M}$ and $\mathcal{P}_{i}$ are the set of all users
and the set of strategy profiles of each user in
(\ref{Powerstratgy}), respectively. The difference between these two games comes from the uncertainty in the utility function of $\widetilde{\mathcal{G}}$, and when $\varepsilon_i=0$ for all players, these two games are the same. In this case, the strategy
profile $\widetilde{\mathbf{p}}^{*}=\{\widetilde{\mathbf{p}}^{*}_{_{1}},\cdots,\widetilde{\mathbf{p}}^{*}_{_{M}}\}$ is RNE of the game
$\widetilde{\mathcal{G}}$ if
\begin{equation}\label{NEpoint}
 \widetilde{u}_{i}(\widetilde{\mathbf{p}}^{*}_{_{i}},\widetilde{\mathbf{p}}^{*}_{_{-i}},\tilde{\mathbf{s}}_i)\geq
 \widetilde{u}_{i}(\mathbf{p}_{_{i}},\widetilde{\mathbf{p}}^{*}_{_{-i}},\tilde{\mathbf{s}}_i),\; \forall \mathbf{p}_{_{i}} \in \mathcal{P}_{i}, \;\forall \tilde{\mathbf{s}}_{i}\in \mathcal{R}_{s}, \; \forall i \in \mathcal{M}.
\end{equation}
meaning that no user can achieve a higher utility by unilaterally changeling its strategy under any condition of uncertainty at RNE.

\section{Analysis of RNE}

We now present our analysis on RNE's existence and uniqueness. By considering uncertainty in the game, equilibrium analysis for $\widetilde{\mathcal{G}}$ becomes more complicated than that for $\mathcal{G}$. This is because the strategy of each user depends on strategies of other users as well as on users' uncertainty regions. In addition, by considering uncertainty in the system, a new coupling is introduced in the game, which requires new signalling between users \cite{Robustnew}. Hence, design and implementation of such a network is more complicated when RNE is considered. Comparing the total achieved utility of the system (i.e., the social utility) at RNE with that at NE provides a measure of performance when robustness is introduced. Considering these issues, we wish to
\begin{itemize}
  \item Simplify the robust game to the extent possible, with a view to simplifying the implementation of the system, and
  \item Determine the relationship between the social utility at RNE and at NE in both cases of unique NE and multiple NEs, and obtain the relation between users' strategies at RNE and the bound on uncertainty.
\end{itemize}

\subsection{Existence and Uniqueness of RNE}

For any realization of error in the uncertainty region, the utility function for $\widetilde{\mathcal{G}}$ in the worst-case robust approach, denoted by $\psi_i(\mathbf{p}_{_{i}},\mathbf{p}_{_{-i}})$ is
\begin{equation}\label{rho}
    \psi_i(\mathbf{p}_{_{i}},\mathbf{p}_{_{-i}})\triangleq \min_{\tilde{\mathbf{s}}_{i}\in \mathcal{R}_{s}
    }\sum_{k=1}^{k}\log(1+\frac{p^{k}_{i}}{\tilde{s}^{k}_{i}}), \quad
    \forall i \in \mathcal{M}.
\end{equation}
Since the norm function is convex \cite{boydconvexbook}, and $\mathcal{R}_{s}$ in (\ref{uncertaintyregion1}) is assumed to be bounded and convex,
$\psi_i(\mathbf{p}_{_{i}},\mathbf{p}_{_{-i}})$ is continuous, and is concave on $\mathbf{p}_{_{i}}$, when $\mathbf{p}_{_{-i}}$ is fixed. Therefore, for any channel realization, any bound on the transmit power, and any constraint on spectral mask, there is an equilibrium, called RNE, for the game $\widetilde{\mathcal{G}}$ (Theorem 2 in \cite{Robustgame}).

Although one can establish the existence of RNE from the characteristics of $\mathcal{R}_s$, obtaining RNE requires excessive calculations and depends on the representation of the uncertainty region, meaning that the optimal transmit power cannot be obtained in a closed form. Hence, the conditions for RNE's uniqueness cannot be obtained in general by the fixed point approach and contraction mapping as in \cite{Nash1,Nash2}.

In practice, uncertainties in sub-channels can be considered as independent and identically distributed (iid) random variables with Gaussian distribution, e.g., noise in wireless channels  \cite{DRO1,DRO2,Bambos2000}, the uncertainty region can be modeled \cite{Robustgame} as
\begin{equation}\label{Uncertainityregionlinear}
\hat{s}^{k}_{i}=[-\varepsilon^{k}_{i}\bar{s}^{k}_{i},\varepsilon^{k}_{i}
\bar{s}^{k}_{i}],
\end{equation}
where $\varepsilon^{k}_{i}>0$. Compared to (\ref{uncertaintyregion1}), the uncertainty region defined by (\ref{Uncertainityregionlinear}) is decomposed into independent intervals for each sub-channel. Hence, a bound on uncertainty, denoted by $\varepsilon^{k}_{i}$ is considered for each sub-channel and each user, which indicates the absolute distance between the nominal and the actual interference levels in each sub-channel. For this type of uncertainties, we can simplify the game $\widetilde{\mathcal{G}}$, and derive the conditions for uniqueness of its RNE.

\textbf{Proposition 1.} When (\ref{Uncertainityregionlinear}) holds, RNE of the robust game $\widetilde{\mathcal{G}}$ is the same as NE of the game $\mathcal{G}$ with the same number of users and strategy profile, and the optimal strategy is the solution to the following problem
\begin{equation}\label{utilitywithuncertainties}
\max_{\mathbf{p}_{_{i}}, \mathbf{p}_{_{-i}}}
\sum_{k=1}^{k}\log(1+\frac{p^{k}_{i}}{\bar{s}^{k}_{i}(1+\varepsilon^{k}_{i})})
\end{equation}
\[\mathrm{subject~to} \left\{
\begin{array}{l l}
C_1: \sum_{k=1}^{K} p^{k}_{i} \leq  p_{i}^{\scriptsize{\textnormal{max}}} \nonumber \\
C_2: p^{k}_{i}\leq  p^{k}_{\scriptsize{\textnormal{mask}}}. \nonumber \\
\end{array} \right. \]
\begin{proof}
See Appendix B.
\end{proof}
By using Proposition 1, the uncertain value of $s_{i}^k$ in the utility function of each user is replaced by the scalar variable $(1+\varepsilon^{k}_{i})$ multiplied by the estimate of $\bar{s}_{i}^k$. In this way, the uncertainty region is represented in a deterministic manner in the robust optimization formula. Comparing the game $\widetilde{\mathcal{G}}$ whose utility is stated in (\ref{IWFrobustcounterpart}) with the robust game $\widetilde{\mathcal{G}}$ whose utility is stated in (\ref{utilitywithuncertainties}) shows that there is no new optimization variable in (\ref{utilitywithuncertainties}). As such, analysis of the robust game $\widetilde{\mathcal{G}}$ is simplified and is similar to that of the conventional IWFA. The solution to (\ref{utilitywithuncertainties}) can be obtained similar to that of (\ref{IWFutility1}) by utilizing the Lagrange dual function as
\begin{equation}\label{optimalopt1robustgame}
    p_{i}^{k}=[\frac{1}{\lambda_{i}}-\bar{s}_{i}^{k}(1+\varepsilon_{i}^{k})]_{0}^{p^{k}_{\scriptsize{\textnormal{mask}}}},
\end{equation}
where $\lambda_i$ is the nonnegative Lagrange multiplier that satisfies
(\ref{optimaloptlambda}). Next, we derive the condition for uniqueness of RNE in the robust game using the framework in \cite{Nash1}.

\textbf{Proposition 2.} When (\ref{Uncertainityregionlinear}) holds, RNE is unique if
\begin{equation}\label{Proposition 2.1}
   \!\! \min\{\frac{\rho(\overline{\textbf{W}}(k)+\overline{\textbf{W}}^{T}(k))}{2},\|\overline{\textbf{W}}(k)\|_2\}+\|\mathbf{w}(k)\|_2<1, \quad \forall k \in \mathcal{K},
\end{equation}
where $\rho$ is the spectral radius of the matrix, $\overline{\textbf{W}}(k)$ is a $M \times M$ matrix whose elements are
\begin{eqnarray}\label{Proposition 2.3}
 \overline{W}_{ij}(k)\triangleq \left\{\begin{array}{l l}
0 \qquad\qquad \textnormal{if}
\qquad i=j \\
 \frac{h^{k}_{ji}}{h^{k}_{ii}} \qquad\quad\, \textnormal{if} \qquad i\neq j, \end{array} \right.
 \end{eqnarray}
the value of $\|\overline{\textbf{W}}(k)\|_2$ is the $l_2$-norm of $\overline{\textbf{W}}(k)$, and $\mathbf{w}(k)=[\varepsilon^{k}_1,\cdots,\varepsilon^{k}_M]$. When $\overline{\textbf{W}}(k)$ is symmetric, (\ref{Proposition 2.1}) reduces to
\begin{equation}\label{Proposition 2.2}
    \rho(\overline{\textbf{W}}(k))+\|\mathbf{w}(k)\|_2<1, \qquad k \in \mathcal{K}.
\end{equation}
\begin{proof}
See Appendix C.
\end{proof}
The difference between the condition for uniqueness of RNE in the robust game obtained by (\ref{Proposition 2.2}) and that of the nominal game comes from $\|\mathbf{w}(k)\|_2$, which indicates that the condition for RNE's uniqueness is tighter than that of the nominal game. This means that for some values of direct and interference channel gains in the robust game, (\ref{Proposition 2.2}) is not satisfied, but the same values satisfy the condition for uniqueness of NE in the nominal game, i.e., when $\varepsilon_i^k=0$. Hence, enlarging the uncertainty region of the system parameters reduces the probability of having a unique RNE in the robust game as compared to the nominal game.

\subsection{Comparison of Social Utility at RNE and at NE}

Now we discuss the effect of uncertainty on RNE of $\widetilde{\mathcal{G}}$ as compared to NE of $\mathcal{G}$ in
terms of the total throughput of users and the number of sub-channels utilized by each user.

\textbf{Theorem 1.} When Proposition 2 holds,
\begin{enumerate}
 \item The strategy of the robust game $\widetilde{\mathcal{G}}$, denoted by $\widetilde{\mathbf{p}}^*$, is a decreasing function of $\varepsilon^k_i$, i.e.,
      \begin{equation}\label{theroem1-1}
   \exists \, \varepsilon_i^{1k} <  \varepsilon_i^{2k} \Rightarrow \widetilde{\mathbf{p}}^{*1} \geq \widetilde{\mathbf{p}}^{*2}
\end{equation}
where  $\widetilde{\mathbf{p}}^{*1}$ and $\widetilde{\mathbf{p}}^{*2}$ are the strategy at RNE for $\varepsilon_i^{1k}$ and $\varepsilon_i^{2k}$, respectively.
\item The social utility at RNE of $\widetilde{\mathcal{G}}$ is less than that of $\mathcal{G}$, and is a decreasing function of $\varepsilon^k_i$, i.e.,
      \begin{equation}\label{theroem1-1}
   \exists \, \varepsilon_i^{1k} <  \varepsilon_i^{2k} \Rightarrow \widetilde{\mathbf{u}}^{*1} \geq \widetilde{\mathbf{u}}^{*2}
\end{equation}
where $\widetilde{\mathbf{u}}^{*1}$ and $\widetilde{\mathbf{u}}^{*2}$ are the social utility for $\varepsilon_i^{1k}$ and $\varepsilon_i^{2k}$, respectively.
 \end{enumerate}
\begin{proof}
See Appendix D.
\end{proof}

From Theorem 1, uncertainty will definitely reduce the total throughput of users when RNE is unique, and the strategy at RNE is a decreasing function of $\varepsilon_i^k$ for all users. However, this may not be true when interference is high, i.e., when we encounter multiple NEs in $\mathcal{G}$. By considering robustness in such cases, from (\ref{optimalopt1robustgame}), we see that users with higher values of $\varepsilon_{i}^{k}$ have to reduce their transmit power in those sub-channels compared to users with smaller values of  $\varepsilon_{i}^{k}$. As such, they cause less interference to other users, which is advantageous from other users' points of view. Consequently, we may see a higher total throughput at a particular RNE as compared to the case for the corresponding NE, depending on the interfering and direct channel gains between users \cite{ProbabilisticIWFA,Robustnew}. To explain the effect of uncertainty in the case of multiple NEs, we define the orthogonal equilibrium in interference channels. At orthogonal equilibrium, the transmit power levels of different users over each sub-channel are non-overlapping. Consider a subset of sub-channels denoted by $\mathcal{K}_i\subseteq \mathcal{K}$ utilized by user $i$. At orthogonal equilibrium, we have
\begin{equation}\label{orthogonal}
    \mathcal{K}_i \cap \mathcal{K}_j=\emptyset \quad \forall i , j \in \mathcal{M},
\end{equation}

\textbf{Proposition 3.} Uncertainty in the game $\widetilde{\mathcal{G}}$ causes convergence to a RNE that has more orthogonality than at NE of the game $\mathcal{G}$.

\begin{proof}
See Appendix E.
\end{proof}

Note that when interference is high, which corresponds to multiple NEs, the IWFA is suboptimal, but orthogonal power allocation is optimal \cite{SimultanousWFA,spectrumsharing}. By considering uncertainty, users are forced to the orthogonal power allocation. Hence, the total throughput at RNE may be higher than that of the corresponding NE, depending on the values of uncertain parameters, i.e., the interfering and direct channel gains between users, noise levels, and power limitations \cite{ProbabilisticIWFA,Robustnew}. When multiple NEs occur in the game, the social utility of users is a non-smooth and non-convex function, and hence, in general, one cannot state that its value at RNE is higher or lower than its corresponding value at NE of the nominal game.

\section{Distributed Algorithm}

A distributed power allocation algorithm without any message passing between users is very desirable for spectrum sharing in unlicensed bands. For conventional IWFA, different iterative algorithms are presented in \cite{scutariasynch,Nash2,Yu} that include simultaneous, sequential and asynchronous updating procedures. All existing algorithms pertaining to conventional IWFA can be applied in our robust approach without any additional complexity, because uncertainty in our formulation is replaced by a scalar value multiplied by the estimated value of the uncertain parameter. Among them, we focus on the asynchronous algorithm because each user maximizes its own utility in a totally asynchronous manner, and there is no need for any coordination or procedure between users for updating the allocated transmit power levels. As such, it is the preferred approach in spectrum sharing environments. Before implementing the asynchronous algorithm for the robust approach, we fist present an overview of the assumptions in the asynchronous iterative waterfilling scheme \cite{scutariasynch}.

We assume that users update their transmit power levels at discrete instances denoted by $\mathcal{T}=\{0,1,2,\cdots\}$, and $\mathbf{p}_i(t)$ denotes the transmit power level of user $i$ at iteration $t$. During $t \nsubseteq \mathcal{T}_i$, the transmit power level of user $i$ is unchanged. Let $T$ be the total number of iterations for users' updating of their transmit power levels, and $t_i^{s}$ be the last time that user $i$ measured the interference from other users at iteration $t$, where $0<t_i^{s}<t$. We denote the vector of measured interference levels over $K$ sub-channels at time $t_i^{s}$ by $\mathbf{s}_{i}(t_i^{s})=[s_{i}^1(t_i^{s}),\cdots,s_{i}^K(t_i^{s})]$. At each iteration $t$ when user $i$ updates its transmit power level, it uses  $\mathbf{s}_{i}(t_i^{s})$ to solve its optimization problem. The asynchronous distributed algorithm is summarized in Table \ref{Table1}.

\begin{table}[ht]\caption{Distributed Algorithm} \label{Table1}
\centering
\begin{tabular}{l}
\hline
\textbf{Asynchronous Worst-Case IWFA}\\
\hline \hline
 \textbf{Inputs for each user}
\\  $\mathcal{T}_i$: The set of iteration times of user $i$
\\ $t_i^{s}$: Last instance of measuring interference from other users for each iteration,
\\ $\mathbf{s}_{i}(t_i^{s})$: Measured interference by user $i$ at $t_i^{s}$,
\\ $\varepsilon_i^k$: Uncertainty region for all users in all sub-channels
\\ \textbf{Initialization} For $t=0$, set any feasible power allocation $\textbf{p}_i(0) $ for all $i \in \mathcal{M}$,
\\ \textbf{Iterative algorithm}   For $t=1,2,\cdots,T$:\\   %
Update the transmit power of each user by \\
 $\mathbf{p}_i(t+1) =  \left\{\begin{array}{l l}
   \text{the solution of~} (\ref{optimalopt1robustgame}) \text{~based on }  \mathbf{s}_{i}(t_i^{s}) \quad\; \text{when} \qquad t \in \tau_i  \nonumber \\
  \mathbf{p}_i(t) \qquad\qquad\!\!\!\!\!  \text{otherwise} \nonumber \\\end{array}, \right.
 $
\\ End.
\\ \hline
\end{tabular}
\end{table}

An important issue for any iterative algorithm is to determine the conditions for its convergence. We derive the condition for the convergence of the asynchronous distributed algorithm to its RNE in Proposition 4 using the sufficient condition for convergence of the conventional asynchronous IWFA \cite{scutariasynch}.

\textbf{Proposition 4.} As $T \rightarrow \infty$, the asynchronous distributed algorithm for the robust IWFA converges to the unique RNE from any initial power allocation $\textbf{p}_i(0)$ if
$\|\mathbf{\overline{W}}^{\scriptsize{\textnormal{max}}}\|_{2}+\sqrt{|M|}\|\mathbf{w}^{\scriptsize{\textnormal{max}}}\|_{2}
< 1$,
where $\mathbf{\overline{W}}^{\scriptsize{\textnormal{max}}}$ is a $M \times M$ matrix whose elements are
\begin{eqnarray}\label{TT}
 \overline{W}_{ij}^{\scriptsize{\textnormal{max}}} \triangleq  \left\{\begin{array}{l l}
0 \qquad\qquad\qquad\,\; \textnormal{if}\qquad i=j \\
 \max_{k \in \mathcal{K}}\frac{h^{k}_{ji}}{h^{k}_{ii}} \qquad \textnormal{if} \qquad i\neq j, \end{array} \right.
 \end{eqnarray}
 and $\mathbf{w}^{\scriptsize{\textnormal{max}}}$ is a $M \times 1$ vector whose $i^{\scriptsize{\textnormal{th}}}$ element is $\max_{k \in \mathcal{K}} \bar{s}_{i}^{k}\varepsilon_{i}^{k} $
\begin{proof}
See Appendix F.
\end{proof}

From Proposition 4, the condition for convergence of the asynchronous algorithm at RNE depends on the uncertainty region as well as on the number of users in the system. A larger uncertainty region and/or a higher number of users reduces the probability of convergence of the asynchronous algorithm to its optimal point. Again, we see that this condition is tighter than that of the game without uncertainty, meaning that the probability of divergence of the distributed algorithm at RNE is higher than that of the distributed algorithm at NE.

\section{Simulation Results}

Now we provide simulation results to get an insight into the performance of $\widetilde{\mathcal{G}}$ for different bounds on uncertainty as compared to $\mathcal{G}$. In the following simulation, for convenience we assume that the values of $\varepsilon_{i}^k$ are equal for all users and all sub-channels, and denote it by $\varepsilon$. In all simulations, we compare the total utility of users at RNE, which is the Shannon rates for users.

\subsection{Unique NE}
\begin{figure}
\centering
\includegraphics [height=6.5cm,width=11.5cm] {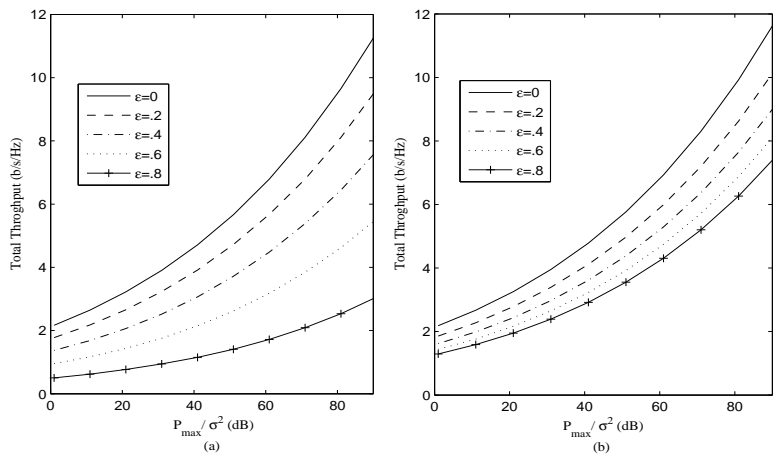}
\caption{Total throughput of users for different values of
$\varepsilon$ when Proposition 2 holds (i.e., when multiuser interference is low).}{\label{fig1}}
\end{figure}

Figs. \ref{fig1}(a) and \ref{fig1}(b) show the effect of
uncertainty on the total throughput of users when Proposition 2  holds, i.e., when RNE is unique, and multiuser interference is low. In this set up, the number of users is 8, the number of sub-channels is 64, and
$p^{k}_{\scriptsize{\textnormal{mask}}}=p^{\scriptsize{\textnormal{max}}}$.
The values of $h^{k}_{ii}$, $h^{k}_{ji}$, and $\sigma^{k}_i$ are randomly chosen from the intervals $[0, 0.1]$, $[0, 0.01]$, and
$[0,0.01]$, respectively, guaranteing that Proposition 2 holds,
and are multiplied by fading coefficients. The estimated error is assumed to be symmetrically distributed in $[-\varepsilon,
\varepsilon]$ for all 64 sub-channels and for all users, and is added to the nominal value of $s_{i}^{k}$.

In Fig. \ref{fig1}(a), the value of $s_{i}^{k}$ is uncertain, but in Fig. \ref{fig1}(b), while the exact value of $s_{i}^{k}$ is available, it is assumed to be uncertain, i.e., $s_{i}^{k}$ is replaced by $(1+\varepsilon)s_{i}^{k}$. We take the average of total throughput values of all users for 20 realizations of channel gains, each with a different error value $\varepsilon$. Note that expanding the bound on uncertainty, reduces the total throughput of users as compared to the case that there is no uncertainty, as expected form Theorem 1, since the social utility of the robust game is a decreasing function of $\varepsilon$. The impact of uncertainty in Fig. \ref{fig1}(b) is much less than that in Fig. \ref{fig1}(a), meaning that in such cases, the utility of each user is not significantly affected by higher interference.

\begin{figure}
\centering
\includegraphics [height=6.5cm,width=11.5cm] {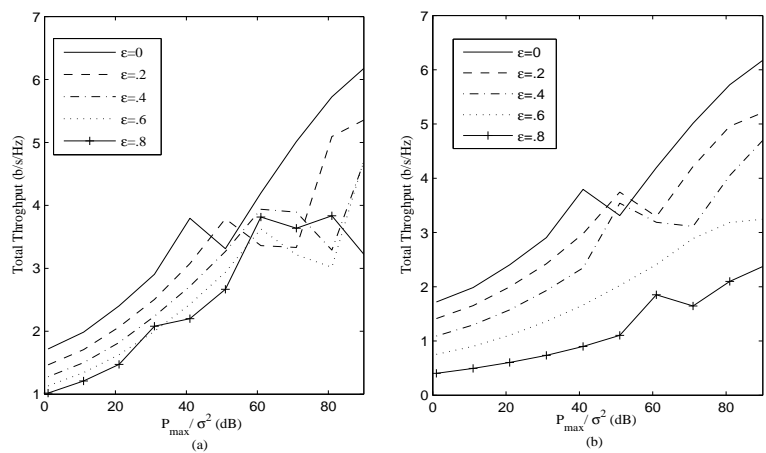}
\caption{Total throughput of users for different values of
$\varepsilon$ when Proposition 2 does not hold (i.e., when multiuser interference is high).}{\label{fig2}}
\end{figure}

\subsection{Multiple NEs}
To show the impact of uncertainty in cases that we encounter
multiple NEs, we assume that Proposition 2 does not hold, meaning that multiuser interference is very high. The values of $h^{k}_{ii}$, $h^{k}_{ji}$, and $\sigma^{k}_i$ are randomly chosen from the intervals $[0, 0.1]$, $[0, 1]$, and $[0,0.01]$ respectively. Again, we consider 8 users and 64 sub-channels, and take the average of total throughput values of all users for 20 realizations of channel gains, each with a different error value $\varepsilon$. The results are shown in Figs. \ref{fig2} (a) and (b). In Fig. \ref{fig2} (a), users encounter  uncertainty, and in Fig. \ref{fig2} (b), although the exact value of $s_{i}^{k}$ is available, it is assumed to be uncertain, i.e., $s_{i}^{k}$ is replaced by $(1+\varepsilon)s_{i}^{k}$.

As expected from Proposition 3, for some values of uncertainty, the robust game has a better performance than that of the conventional IWFA in both of the above cases, because uncertainty causes the game to converge to the orthogonal NE, resulting in a higher total throughput in some cases. The same is numerically shown in \cite{Robusthaykin}, where the average throughput in the robust power allocation problem is shown to be very close to that of the conventional IWFA, and in \cite{ProbabilisticIWFA}, where a higher throughput for the probabilistic robust algorithm is reported as compared to that of the case where complete information is utilized by IWFA. Also, as we see in Fig. \ref{fig2} (b), in high interference scenarios, a higher estimate for the interference is advantageous as it may lead to orthogonal power allocations and consequently, a higher utility for each user, as expected from Proposition 3.

\begin{table}[h]
\caption{Channel Gains of 3 users} \vspace{-.2in}
\centering
\begin{tabular}{|c|c|c|c|c|c|c|}
  \hline
  k & 1 & 2 & 3 & 4& 5 & 6 \\
  \hline
  $h_{11}$ & 20.52 &  2.0  &  2.08 &  10.56  &  0.44 &   1.6 \\
$h_{12}$ &   4.91  &  4.97 &   3.95 &   3.94& 2.95& 5.95\\
$h_{13}$ &  7.9  &  5.97 &   2.97  &  4.92  &  1.93 & 6.94\\
$h_{21}$ & 0.92  & 0.94  &  0.95 &   0.92   & 0.95& 0.99\\
$h_{22}$ & 2.44&  26.32&   23.2&    3.64&    3.92& 0.68 \\
$h_{23}$ & 0.91  &  0.96   & 0.99   & 0.99  &  0.934 &0.95\\
$h_{31}$ & 0.91 &  0.95&    0.98&    0.98   & 0.93& 0.96\\
$h_{32}$ & 0.93&    0.96&    0.90&    0.96&    0.98&    0.97\\
$h_{33}$&  3.6&   24&    6&    1.6&   34& 40\\
$\sigma^{2}_{1}$ &2.2&    0.26&    4.1&    3.06&    0.02& 0.02\\
$\sigma^{2}_{2}$ &    8.24&    0.08&    0.18&    0.08& 0.04&0.06\\
  $\sigma^{2}_{3}$ &  0.22&    0.26&    4.08&    1.06&    0.02&
  0.02\\
  \hline
\end{tabular}
\label{tabel1}
\end{table}

To observe how the robust power allocation algorithms converge to the orthogonal equilibrium, let us consider 3 users in the system, $K=6$, $p^{\scriptsize{\textnormal{max}}}=1$, $p^{k}_{\scriptsize{\textnormal{mask}}}=0.5$ Watts, and channel gains and noise coefficients as in Table \ref{tabel1}. In Tables \ref{tabel2} and \ref{tabel3}, we show the allocated power to each user in $\mathcal{G}$ and $\widetilde{\mathcal{G}}$, respectively. As can be seen for $\mathcal{G}$, users spread their allocated transmit power in different sub-channels, and induce interference to other users. Note that user 1 uses $k={1,2,4}$, whereas at RNE of $\widetilde{\mathcal{G}}$, each user concentrates its power on the best sub-channels, e.g., user 1 only uses sub-channels 1 and 4, whose direct channel gains denoted by $h_{11}^{1}$ and $h_{11}^{4}$ are higher than those of other sub-channels, and whose interference channel gains denoted by $h_{21}^{1}$, $h_{21}^{4}$, $h_{31}^{1}$, and $h_{31}^{4}$ are lower than  those of other sub-channels as shown in Table \ref{tabel1}. The same is true for users 2 and 3. This means that in the game $\widetilde{\mathcal{G}}$, all users use their best sub-channels and the allocated transmit power levels to users are orthogonal, i.e., there is no interference between users; hence the total aggregate throughput of users is increased. As can be seen in Tables \ref{tabel2} and \ref{tabel3}, in this case, the throughput of each user is also increased as compared to those of the game $\mathcal{G}$. However, in general, the throughput of each user depends on its channel, and may not be increased in all instances in the robust game.

\begin{table}[h]
\caption{Power Allocation by 3 users at NE of $\mathcal{G}$} \vspace{-.2in}
\centering
\begin{tabular}{|c|c|c|c|c|c|c|c|}
\hline
 $\text{user}_i$ & $u_{i}$ & $p_1$ & $p_2$ & $p_3$ & $p_4$& $p_5$ & $p_6$\\
  \hline
  $\text{user}_1$ &1.92&  0.44  &  0.1&     0 &   0.45 &  0  &       0  \\
$\text{user}_2$& 3.82& 0   & 0.5&    0.5&     0  &       0 &        0 \\
$\text{user}_3$ &  10.9 &0  &  0.0059    &0.3049  &   0  &  0.32 &   0.37 \\
\hline
\end{tabular}
\label{tabel2}
\end{table}

\begin{table}[h]
\caption{Power Allocation by 3 users at RNE of $\widetilde{\mathcal{G}}$ for $\varepsilon=3$} \vspace{-.2in}
\centering
\begin{tabular}{|c|c|c|c|c|c|c|c|}
  \hline
 $\text{user}_{i}$ & $u_{i}$ & $p_1$ & $p_2$ & $p_3$ & $p_4$& $p_5$ & $p_6$  \\
  \hline
  $\text{user}_1$ & 1.93 &  0.5&   0 & 0&    0.5&  0 &  0 \\
$\text{user}_2$ & 3.95&0 &  0.5&    0.5&         0   &      0    &     0  \\
$\text{user}_3$ &  11.17 & 0     &    0   &      0    &     0  &  0.5&    0.5 \\
\hline
\end{tabular}
\label{tabel3}
\end{table}

\begin{figure}
\centering
\includegraphics [height=7cm,width=9.5cm] {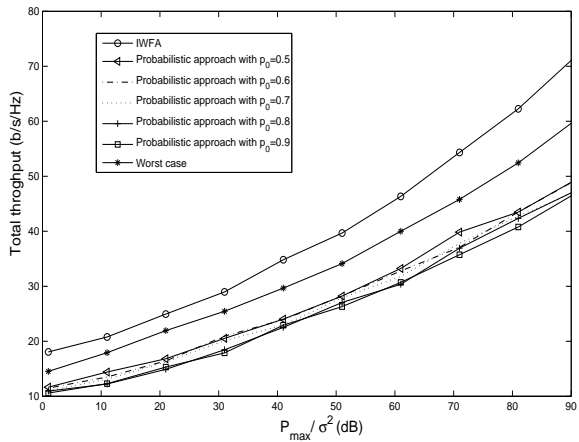}
\caption{Total throughput of users for different values of $\delta_0$ in the probabilistic approach as compared to that of the worst-case approach when Proposition 2 holds (i.e., when multiuser interference is low).}{\label{withprobabilityuniquenash1}}
\end{figure}

\subsection{Comparison of Performances of Worst-Case and Probabilistic Approaches}

\begin{figure}
\centering
\includegraphics [height=7cm,width=9.5cm] {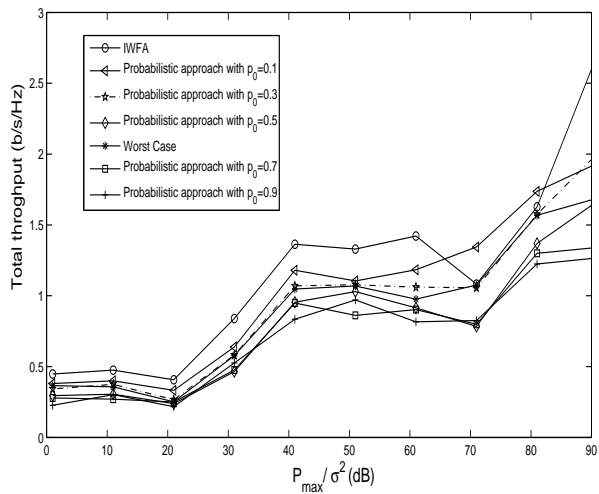}
\caption{Total throughput of users for different values of $\delta_0$ for the probabilistic approach as compared to that of the worst-case approach when Proposition 2 does not hold (i.e., when multiuser interference is high).}{\label{withprobabilitymultiplenash3}}
\end{figure}

Finally, we compare the performance of our proposed RNE with that of the probabilistic approach in \cite{ProbabilisticIWFA} that guarantees that the achieved throughput of users is higher than a specific value with a given probability $\delta_0$ under any level of uncertainty in the value of IN in each sub-channel. The value of $\delta_0=0$ means that there is no guarantee to achieve the specific data rate under any level of uncertainty in IN, and higher values of $\delta_0$ provides more protection for the required data rate against uncertainty in IN levels. In \cite{ProbabilisticIWFA}, the objective function of the robust game is changed to the probability of achieving different data rates, which is generally hard to solve. To simplify the problem and obtain a closed form solution, they assumed that uncertainty in each parameter is confined to (\ref{Uncertainityregionlinear}) and its probability density function (pdf) is uniform. Under such assumptions, the robust probabilistic counterpart of $\mathcal{G}$ is simplified in such a way that the uncertainty multiplier for $\bar{s}_{i}^{k}$ in (\ref{utilitywithuncertainties}) in our worst-case algorithm, i.e., $(1+\varepsilon)$ is replaced by $(1-\varepsilon+2\times \varepsilon \times \delta_0)$. Hence, one can consider the probabilistic approach as a special case of worst-case approach with uncertainty bound $(1-\varepsilon+2\times \varepsilon \times \delta_0)$. When $0 \leq \delta_0 < 0.5$, the effect of uncertainty on NE of the probabilistic approach is less than that of the worst-case approach; and when $0.5 \leq \delta_0 \leq 1$, the probabilistic approach is more robust than the worst-case approach. This phenomenon is shown in Figs. \ref{withprobabilityuniquenash1} and \ref{withprobabilitymultiplenash3} for the case with a unique RNE and for multiple RNEs, respectively.

As another example, we now consider $\varepsilon=0.8$ to compare the performances of the worst-case and probabilistic approaches for different values of $\delta_0$. When RNE is unique and $0 < \delta_0 < 0.5$, i.e., $(1-\varepsilon+2\times \varepsilon \times \delta_0)<(1+\varepsilon)$, the social utility of the probabilistic approach is higher than that of the worst-case approach, which is in line with Theorem 1, since the social utility is a decreasing function of the size of the uncertainty region. In contrast, for $0.5 \leq \delta_0 \leq 1$, i.e., when $(1-\varepsilon+2\times \varepsilon \times \delta_0) \geq (1+\varepsilon)$, the social utility of the worst-case approach is higher than that of the probabilistic approach at the RNE. This may not be true for the case of multiple RNEs. As stated in Proposition 3, uncertainty may result in more orthogonality at a RNE as compared to the non-robust approach and/or to the case with less uncertainty. In such cases, the total throughput may be higher at a RNE, as shown in Fig. \ref{withprobabilitymultiplenash3}.

\section{Conclusions}

We studied the impact of uncertainty in the channel state information and in interference levels on the total throughput of users in shared unlicensed spectrum via robust game theory. To devise a robust power allocation in such environments, we focused on bounded and symmetrically distributed uncertainties, and showed that this game can be considered as a conventional IWFA with system parameters multiplied by bounds on uncertainty. The conditions for existence and uniqueness of RNE and for convergence of the proposed distributed algorithms of the robust game were derived via the conventional IWFA. The performance of the robust game was compared to that of the conventional IWFA in terms of its total throughput. In the case of multiple RNEs, we showed that the orthogonal use of sub-channels by users at a RNE may lead to a higher total throughput of users as compared to that of the conventional IWFA.

\appendices

\section{Modeling Uncertainty by Norm Function}

In this appendix, we explain the relationship between the size of the uncertainty region denoted by $\varepsilon_i$, and the stochastic nature of the uncertain parameters by utilizing Section 8.5.5. in \cite{Gershman} and \cite{Robustcogref17}. Let $\mathbf{x}$ be the uncertain parameter for which we have $\hat{\mathbf{x}}=\tilde{\mathbf{x}}-\bar{\mathbf{x}}$, where $\tilde{\mathbf{x}}$, $\bar{\mathbf{x}}$, and $\hat{\mathbf{x}}$ are the exact value, the nominal (or estimated) value, and the error in $\mathbf{x}$, respectively. Assume that error has a specific probability distribution function (pdf) $f(\hat{\mathbf{x}})$, such as Gaussian or uniform. One can find the uncertainty region in such a way that with probability $P$, any realization of uncertain parameter falls in the uncertainty region. The generic formulation of this statement is
\begin{equation}\label{pdf}
    \int_{\|\tilde{\mathbf{x}}-\bar{\mathbf{x}}\|\leq \varepsilon_i} f(\tilde{\mathbf{x}}-\bar{\mathbf{x}}) d \hat{\mathbf{x}}\leq P
\end{equation}
For different types of uncertainty, each with a different pdf, a different uncertainty region can be obtained. 

\section{Proof of Proposition 1}

As in Theorem 4 in \cite{Robustgame}, consider a robust finite game, with $M$ players, without private information, in which $u_{i}(\textbf{x}_{i},\textbf{a}_{i})$ is a continuous and differentiable utility function for the $i^{{\scriptsize{\textnormal{th}}}}$ player, $\textbf{x}_{i}$ is the strategy vector of that player, $\textbf{a}_{i}$ is the vector of uncertain parameters where $a_{ji}$ is its
$j^{{\scriptsize{\textnormal{th}}}}$ element, and its uncertainty region is defined by the following interval
\begin{equation}
    a_{ji}=[a_{ji}^{\scriptsize{\textnormal{lower}}},a_{ji}^{\scriptsize{\textnormal{upper}}}], \nonumber
\end{equation}
where $a_{ji}^{\scriptsize{\textnormal{lower}}}$ and
$a_{ji}^{\scriptsize{\textnormal{upper}}}$ are the corresponding lower and upper bounds of the uncertain parameter, respectively. We denote the utility of the nominal game for user $i$ by $u_{i}(\textbf{x}_{i},\textbf{a}^\text{d}_{i})$, where $\textbf{a}^\text{d}_i$ is the deterministic vector of system parameter and $a^\text{d}_{ji}$ is its $j^{{\scriptsize{\textnormal{th}}}}$ element. Note that $\textbf{a}^\text{d}_i$ is not the optimization vector in the nominal game, and that the nominal game has only one optimization vector, namely $\textbf{x}_{i}$.

For the robust game, the utility function for user $i$ is denoted by $\widetilde{u}_{i}( \textbf{x}_{i},\textbf{a}_{i})$, where $\textbf{a}_{i}$ is the vector of uncertain parameter, and both $\textbf{a}_{i}$ and $\textbf{x}_{i}$ are the optimization vectors. The RNE of the robust game is $\{\textbf{x}_{1},\cdots,\textbf{x}_{M}\}$ iff it is also the NE of a finite game with complete information, and with the same number of players and the same strategy space, with the following utility function (Theorem 4 in \cite{Robustgame})
   \begin{eqnarray}\label{equationofproposition1}
 \widetilde{u}_{i}(\textbf{x}_{i},\textbf{a}^\text{d}_{i})= u_i(\textbf{x}_{i},\tilde{\textbf{a}}^\text{d}_{i} ),
\end{eqnarray}
where $\tilde{\textbf{a}}^\text{d}_{i}$ is the new vector related to the uncertain parameter and its $j^{\text{th}}$ element $\tilde{ a}^\text{d}_{ji}$ is
\begin{eqnarray}\label{equationofproposition1}
\tilde{ a}^\text{d}_{ji}= \left\{
\begin{array}{l l}
 a_{ji}^{\scriptsize{\textnormal{upper}}},\quad \text{if}  \quad \frac{\partial \widetilde{u}_{i}(\textbf{a}_{i}, \textbf{x}_{i})}{\partial a_{ji}}\geq0\\
  a_{ji}^{\scriptsize{\textnormal{upper}}} \quad \text{if} \quad  \frac{\partial \widetilde{u}_{i}(\textbf{a}_{i}, \textbf{x}_{i})}{\partial a_{ji}}<0.
\end{array} \right.
\end{eqnarray}
In the above reformulation, the robust utility function has only one optimization vector, i.e., $\textbf{x}_{i}$, and the uncertain parameter is replaced by the deterministic value that depends on the derivative of the robust utility function with respect to its uncertain parameter. In our robust game $\widetilde{\mathcal{G}}$, the derivative of (\ref{utilitydefnition}) with respect to the uncertain parameter $\tilde{s}^{k}_i$ is
\begin{equation}\label{deri}
    \frac{\partial \widetilde{u}_{i}}{\partial \tilde{s}^{k}_i}=\frac{-p_i^k}{\tilde{s}_i^k(\tilde{s}_i^k+p_i^k)}.
 \end{equation}
Since $p_i^k\geq 0$ and interference is always a positive value, we have
 \begin{equation}\label{deri}
    \frac{\partial \widetilde{u}_{i}}{\partial \tilde{s}^{k}_i}<0
 \end{equation}
For our robust game, the upper bound of the uncertain parameter in each sub-channel denoted by $a_{ji}^{\scriptsize{\textnormal{upper}}}$ is equal to $\bar{s}_i^k+\varepsilon_i\bar{s}_i^k$. Hence, the RNE can be obtained when
\begin{equation}
    \widetilde{u}_{i}( \mathbf{p},\mathbf{s}_i^\text{d})= u_i(\mathbf{p},\tilde{\mathbf{s}}_i^\text{d} )
\end{equation}
where $\tilde{\mathbf{s}}_i^\text{d}$ is a deterministic vector of the uncertainty $\mathbf{s}_i$, and $\tilde{s}_i^{\text{d}k}$ is its $k^{\text{th}}$ elements where $\tilde{s}_i^{\text{d}k}=\bar{s}_i^k+\varepsilon_i\bar{s}_i^k$. Hence, we have
\begin{equation}\label{utilityrobust}
    \widetilde{u}_{i}( \mathbf{p},\tilde{\mathbf{s}}_i) = (u_i(\mathbf{p},\tilde{\mathbf{s}}_i^\text{d}))_{\tilde{s}_i^{\text{d}k}=\bar{s}_i^k+\varepsilon_i\bar{s}_i^k} = \sum_{k=1}^{K} \log(1+\frac{p^{k}_{i}}{\bar{s}^{k}_{i}(1+\varepsilon^{k}_{i})}).
\end{equation}

\section{Proof of Proposition 2}

For the complete information game $\mathcal{G}$, NE is unique \cite{Nash1} if
\begin{equation}\label{0.Proof of Proposition 2.}
\rho(\textbf{W}(k))<1 \quad \forall k \in \mathcal{K},
\end{equation}
where $\rho$ is the spectral radius of $\textbf{W}(k)$, and
$\textbf{W}(k)$ is a $M \times M$ matrix whose elements are
\begin{eqnarray}\label{1.Proof of Proposition 2.}
 W_{ij}(k) =  \left\{\begin{array}{l l}
0 \qquad\qquad \textnormal{if}\qquad i=j \\
 \frac{h^{k}_{ji}}{h^{k}_{ii}} \qquad \;\quad\;\!\! \textnormal{if} \qquad i\neq j. \end{array} \right.
 \end{eqnarray}
By considering uncertainty in $\mathbf{s}_i$, the robust game $\widetilde{\mathcal{G}}$ has a unique RNE if
\begin{equation}\label{2.Proof of Proposition 2.}
    \max_{\mathbf{s}_i\in \mathcal{R}_{s} \, \forall i \in \mathcal{M}} \rho (\textbf{W}(k))<1.
\end{equation}
If the uncertainty region is modeled by  (\ref{Uncertainityregionlinear}), we have
   $\textbf{W}(k)= \overline{\textbf{W}}(k)+ \widehat{\textbf{W}}(k)$,
where $\overline{\textbf{W}}(k)$ is a $M \times M$ matrix whose elements are \begin{eqnarray}\label{4.Proof of Proposition 2}
 \overline{W}_{ij}(k) =  \left\{\begin{array}{l l}
0 \qquad\qquad \textnormal{if}\qquad i=j \\
 \frac{h^{k}_{ji}}{h^{k}_{ii}} \qquad \;\quad\;\!\! \textnormal{if} \qquad i\neq j, \end{array} \right.
 \end{eqnarray}
and $\widehat{\textbf{W}}(k)$ is a matrix of the uncertain parts of
system parameters in which the sum of its
$i^{{\scriptsize{\textnormal{th}}}}$ row is less than
 $\varepsilon^{k}_{i}$. Recall that Frobenius norm of a matrix is  $\|\overline{W}_{ij}(k)\|_{F}=\sum_{i,j}\overline{W}^{2}_{ij}(k)$ and that  $\|\overline{W}_{ij}(k)\|_{2}\leq\|\overline{W}_{ij}(k)\|_{F}$. Thus
 \begin{eqnarray}\label{5. Proof of Proposition 2}
     \max_{\tilde{\mathbf{s}}_i\in \mathcal{R}_{s} \, \forall i \in \mathcal{M}} \rho (\textbf{W}(k))\leq  \|\overline{W}_{ij}(k)\|_{2}+   \|\widehat{W}_{ij}(k)\|_{2} \leq \|\overline{W}_{ij}(k)\|_{2}+
     \|\widehat{W}_{ij}(k)\|_{F}\leq\|\overline{W}_{ij}(k)\|_{2}+\|\mathbf{w}^{k}\|_2.
\end{eqnarray}
where $\mathbf{w}^{k}=[\varepsilon_1^k,\cdots,\varepsilon_M^k]$.
On the other hand, from Theorem 1.1 in \cite{Matrixanalysis1} we have
\begin{equation}\label{6.Proof of Proposition 2}
 \max_{_{\|\textbf{G}\|_{F} < 1}} \rho (\textbf{F}+\textbf{G})\leq  \rho
 (\frac{\textbf{F}+\textbf{F}^{T}}{2})+1. \nonumber
\end{equation}
Thus
\begin{equation}\label{7.Proof of Proposition 2}
 \max_{_{\|\widehat{\textbf{W}}\|_{F} < \|\mathbf{w}^{k}\|_2}} \rho (\overline{\textbf{W}}+\widehat{\textbf{W}})\leq  \rho
 (\frac{\overline{\textbf{W}}+\overline{\textbf{W}}^{T}}{2})+\|\mathbf{w}^{k}\|_2. \nonumber
\end{equation}
For a symmetric $\textbf{W}$, we have $\widehat{\textbf{W}}=\widehat{\textbf{W}}^\text{T}$,
$\rho(\frac{\overline{\textbf{W}}+\overline{\textbf{W}}^\text{T}}{2})=\rho(\widehat{\textbf{W}})\leq\|\widehat{W}_{ij}(k)\|_{2}$,
and hence, (\ref{Proposition 2.1}) reduces to (\ref{Proposition 2.2}).

\section{Proof of Theorem 1}
To prove this Theorem, we use variational inequalities (VI) to reformulate NE, and derive the relationship between $\mathbf{p}^*$ and $\widetilde{\mathbf{p}}^*$.

Step 1: Consider that the nominal NE is the solution to the following variational inequalities (Proposition 1.4.2 in \cite{PangVI} and \cite{VIintroduction}),
\begin{equation}\label{RNEVI}
    \mathbf{p}^*= \text{Solution to~}(VI(\mathcal{P}, \mathcal{F}))
\end{equation}
where $\mathcal{P}=\prod_{i \in \mathcal{M}}\mathcal{P}_i$, $\mathcal{F}=(\mathcal{F}_i)_{i=1}^{M}$, $\mathcal{F}_i= \boldsymbol{\sigma}_i+\sum_{j=1}^{M}\mathbf{M}_{ij}\mathbf{p}_{j}$, $\boldsymbol{\sigma}_i=(\sigma_{i}^{k}/h_{ii}^{k})_{k=1}^{K}$, and $\mathbf{M}_{ij}=\text{diag}(\frac{h_{ji}^{k}}{h_{ii}^{k}})_{j=1}^{M}$. Considering uncertainty in the parameters can be viewed as perturbation in $\mathcal{F}=(\mathcal{F}_i)_{i=1}^{M}$, which we show by $\widetilde{\mathcal{F}}=(\widetilde{\mathcal{F}}_i)_{i=1}^{M}$, where $\widetilde{\mathcal{F}}_i$ has the same definition as $\mathcal{F}_i$ except that $\widetilde{\boldsymbol{\sigma}}_i=(\sigma_{i}^{k}(1+\varepsilon_{i}^{k})/h_{ii}^{k})_{k=1}^{K}$ and $\mathbf{M}_{ij}=\text{diag}(\frac{h_{ji}^{k}(1+\varepsilon_{i}^{k})}{h_{ii}^{k}})_{j=1}^{M}$. From the above, we consider the solution to $\widetilde{\mathcal{G}}$ as a solution to $VI(\mathcal{P}, \mathcal{F}+q)$, where $q=\max \varepsilon_{i}^{k}\frac{h_{ij}^{k}}{h_{ii}^{k}}$. For the Affine VI, when Proposition 2 holds, since $\mathbf{M}$ is strictly copositive, $\mathcal{F}$ is strongly monotone (\ref{0.Proof of Proposition 2.}).

Step 2: When $\mathcal{F}$ is strongly monotone, the solution to $VI(\mathcal{P}, \mathcal{F}+q)$, i.e., $\varphi(-q)$ is a monotone plus single-valued map (Corollary 2.9.17 in \cite{PangVI}).

\textbf{Statement}: When $\mathcal{F}$ is strongly monotone, $\varphi(-q)$ is a decreasing function of $-q$.
\begin{proof}
Recall that when $\mathcal{F}$ is strongly monotone, $\varphi(-q)$ is monotone (Corollary 2.9.17 in \cite{PangVI}). To prove this statement, we first assume that $\varphi(-q)$ is an increasing monotone function of $-q$, i.e.,
\begin{equation}\label{C1}
    \text{C1:} \quad -q_1<-q_2 \Rightarrow \varphi(-q_1)<\varphi(-q_2), \nonumber
\end{equation}
From the definition of VI for any concave optimization problem (Chapter 1 in \cite{PangVI}), we have
\begin{equation}\label{VIconcave}
    (\mathbf{b}-\varphi(-q))(\mathcal{F}+q)<0, \quad \forall \mathbf{b} \in  \mathcal{P}.
\end{equation}
Since $\varphi(-q_1)$ and $\varphi(-q_2)$ belong to $\mathcal{P}$, we have
\begin{eqnarray}\label{st1}
  (\varphi(-q_1)-\varphi(-q_2))(\mathcal{F}+q_2)<0 \\ \label{st2}
  (\varphi(-q_2)-\varphi(-q_1))(\mathcal{F}+q_1)<0
\end{eqnarray}
We subtract (\ref{st1}) from (\ref{st2}) and write
\begin{equation}\label{step31}
   ( \varphi(-q_1)-\varphi(-q_2))(q_2 - q_1)<0,
\end{equation}
From C1, we have $(\varphi(-q_1)-\varphi(-q_2))<0$. Hence, to hold (\ref{step31}), we should have $q_2 - q_1>0$ or $q_2>q_1$. Clearly, $q_2>q_1$ contradicts with C1. This contradiction shows that $\varphi(-q)$ is a decreasing monotone function of $-q$.
\end{proof}

In our problem, $\varphi(-q)$ is the solution to the robust game and we have $q=\max \varepsilon_{i}^{k}\frac{h_{ji}^{k}}{h_{ii}^{k}}$ for all sub-channels and all users. This means that by increasing uncertainty, the power allocated to each user, i.e., $\varphi(-q)$, is reduced as compared to that of the nominal game. In other words, when $\varepsilon_i^{1k}>\varepsilon_i^{2k}$ for all sub-channels, and $\widetilde{\mathbf{p}}^*_1$ and $\widetilde{\mathbf{p}}^*_2$ are the RNE of $\widetilde{\mathcal{G}}$ for $\varepsilon_i^{1k}$ and $\varepsilon_i^{2k}$, we have
 \begin{equation}\label{111}
   \varepsilon_i^{1k}>\varepsilon_i^{2k} \quad \forall k,  \Longrightarrow  \widetilde{\mathbf{p}}^*_1 <\widetilde{\mathbf{p}}^*_2.
 \end{equation}
Also, for affine VI, the solution set is closed and continuous (Corollary 2.6.4 of \cite{PangVI}). Therefore, the strategy at RNE, i.e., $\widetilde{\mathbf{p}}^*$ is a decreasing function of $\varepsilon_i^k$ (Proof of Part 1 of Theorem 1 in \cite{saeedeh5}). Also, for the nominal game, when $\varepsilon_i^{2k}=0$, we have
 \begin{equation}\label{111-2}
   \varepsilon_i^{k}>0 \quad \forall k,  \Longrightarrow  \widetilde{\mathbf{p}}^* <\mathbf{p}^*.
 \end{equation}

Step 3: Since $\mathcal{F}$ is strongly monotone, the utility function is strictly concave. Hence, for (\ref{111}), we have
\begin{equation}\label{}
\widetilde{\mathbf{p}}^*_1 <\widetilde{\mathbf{p}}^*_2 \Rightarrow  \sum_{i \in \mathcal{M}} \widetilde{u}(\widetilde{\mathbf{p}}^*_1) <\sum_{i \in \mathcal{M}} \widetilde{u}(\widetilde{\mathbf{p}}^*_2),
\end{equation}
where $\sum_{i \in \mathcal{M}} \widetilde{u}(\widetilde{\mathbf{p}}^*_1)$ and $\sum_{i \in \mathcal{M}} \widetilde{u}(\widetilde{\mathbf{p}}^*_2)$ are the social utilities at RNE of $\widetilde{\mathcal{G}}$ for $\varepsilon_i^{1k}$ and $\varepsilon_i^{2k}$, respectively. The social utility of the nominal game is obtained when $\varepsilon_i^k=0$ for all users and for all sub-channels in $\mathcal{G}$. Since the social utility is a decreasing function of $\varepsilon_i^k$, we have
 \begin{equation}\label{}
    \varepsilon_i^{k}>0  \quad \forall k, \Longrightarrow \mathbf{p}^*>\widetilde{\mathbf{p}}^*  \Longrightarrow \sum_{i \in \mathcal{M}} \widetilde{u}_i(\widetilde{\mathbf{p}}^*) <\sum_{i \in \mathcal{M}} u_i(\widetilde{\mathbf{p}}^*),
 \end{equation}
 where  $\sum_{i \in \mathcal{M}} \widetilde{u}_i(\widetilde{\mathbf{p}}^*) $ and $\sum_{i \in \mathcal{M}} u_i(\widetilde{\mathbf{p}}^*)$ are the social utilities of $\mathcal{G}$ and $\widetilde{\mathcal{G}}$, respectively. Therefore, the social utility of the robust game is always less than that of the nominal game.

\section{Proof of Proposition 3}

Assume that the $i^\text{th}$ user converges to $\mathbf{p}_{i}^{*}$ at NE of $\mathcal{G}$, for which there is a corresponding $s_{i}^{k*}$, and the set of all sub-channels with nonzero power allocation by the $i^{\scriptsize{\textnormal{th}}}$ user is denoted by $\mathcal{I}_{i}^{*}$. Consequently,
$\widetilde{\mathbf{p}}_{i}^{*}$,
$\widetilde{s}_{i}^{*k}$, and
$\widetilde{\mathcal{I}}_{i}^{*}$ belongs to the game
$\widetilde{\mathcal{G}}$. In what follows, we will show that
$\widetilde{\mathcal{I}}_{i}^{*}\subseteq \mathcal{I}_{i}^{*}$. In doing so, we denote the Lagrange multipliers at NE of the games
$\mathcal{G}$ and $\widetilde{\mathcal{G}}$ by $\lambda_{i}^{*}$ and $\widetilde{\lambda}_{i}^{*}$, respectively, which are
increasing functions of $s_{i}^{k}$. For user $i$ we have
\begin{eqnarray}  \label{Lemma 2.1}
 \frac{1}{\widetilde{\lambda}_{i}^{*}}\leq \frac{1}{\lambda_{i}^{*}}= \sum_{k \in \mathcal{I}_{i}^{*} }
s_{i}^{k*}+p_{i}^{\scriptsize{\textnormal{max}}} <
\frac{\sum_{i\neq j} p^{q*}_{j}h^{q}_{ji}+\sigma^{q}}{h_{ii}^{q}},
\end{eqnarray}
where $q \notin \widetilde{\mathcal{I}}_{i}$. Obviously, $\mathbf{p}_{i}^{*}$ of the $i^\text{th}$ user leads to $s^{k*}_{i}\leq s^{k}_{i} \quad \forall i , k$. Therefore, for any power allocation strategy $\mathbf{p}_{i} \in \mathbf{p}$, we have
\begin{eqnarray}\label{Lemma 2.2}
\frac{\sum_{i\neq j} p^{q*}_{j}h^{q}_{ji}+\sigma^{q}}{h_{ii}^{q}}
<(1+\varepsilon^{q}_{i})\frac{\sum_{i\neq
j}p^{q}_{j}h^{q}_{ji}+\sigma^{q}}{h_{ii}^{q}}.
\end{eqnarray}
From (\ref{Lemma 2.1}) and (\ref{Lemma 2.2}), those sub-channels that are not used in $\mathcal{G}$ are not used in
$\widetilde{\mathcal{G}}$ as well. Now, as stated in Theorem 1 in \cite{Nash1}, when
multi-user interference is high, an orthogonal NE always exists
for the game $\mathcal{G}$. When the $i^{\scriptsize{\textnormal{th}}}$ user chooses a $q' \in
\widetilde{\mathcal{I}}_{i}$, it is orthogonal to those of other users. Assuming this, for any other power allocation profile of other users, we have
\begin{equation}\label{Lemma 2.3}
(1+\varepsilon^{q'}_{i})\frac{\sigma^{q'}}{h_{ii}^{q'}}<
(1+\varepsilon^{q'}_{i})\frac{\sum_{i\neq j}
p^{q'}_{j}h^{q'}_{ji}+\sigma^{q'}}{h_{ii}^{q'}}, \quad \forall i \in \mathcal{M}.
\end{equation}
Since $\widetilde{s}_{i}^{*k}\gg 1$, the utility at RNE for each user is higher than that of any other strategy profile. Hence, there is no incentive for other users to change their strategy profile from the orthogonal RNE. As such, RNE of $\widetilde{\mathcal{G}}$ is more orthogonal than NE of $\mathcal{G}$.
\section{Proof of Proposition 4}
For the game with complete information $\mathcal{G}$, it is proved in Theorems 3 in \cite{scutariasynch} that the asynchronous distributed algorithm converges if
\begin{equation}\label{1. Proof of Theoream 1.}
\rho(\textbf{W}^{\scriptsize{\textnormal{max}}})<1 \quad \forall k \in \mathcal{K},
\end{equation}
where $\rho$ is the spectral radius of
$\textbf{W}^{\scriptsize{\textnormal{max}}}$, and $\textbf{W}^{\scriptsize{\textnormal{max}}}$ is a $M \times M$ matrix whose elements are
\begin{eqnarray}\label{2. Proof of Theoream 1.}
 W_{ij}^{\scriptsize{\textnormal{max}}}=  \left\{\begin{array}{l l}
0 \qquad\qquad \qquad\,\;\textnormal{if}\qquad i=j \\
\max_{k \in \mathcal{K}} \frac{h^{k}_{ji}}{h^{k}_{ii}} \qquad
\textnormal{if} \qquad i\neq j.
\end{array} \right.
 \end{eqnarray}
From (\ref{Uncertainityregionlinear}), the robust game
$\widetilde{\mathcal{G}}$ converges to RNE by using the iterative asynchronous algorithm if
\begin{equation}\label{3. Proof of Theoream 1.}
    \max_{\tilde{\mathbf{s}}_{i} \in \mathcal{R}_{s} , \forall i, k} \rho (\textbf{W}^{\scriptsize{\textnormal{max}}})<1.
\end{equation}
In this case, we have $\textbf{W}^{\scriptsize{\textnormal{max}}}= \overline{\textbf{W}}^{\scriptsize{\textnormal{max}}}+
   \widehat{\textbf{W}}^{\scriptsize{\textnormal{max}}}$, where $\widehat{\textbf{W}}^{\scriptsize{\textnormal{max}}}$ is the uncertain parts of $\textbf{W}^{\scriptsize{\textnormal{max}}}$, whose elements are
\begin{eqnarray}\label{5. Proof of Theoream 1.}
 \overline{W}_{ij}^{\scriptsize{\textnormal{max}}}=  \left\{\begin{array}{l l}
0 \qquad\qquad \;\;\;\textnormal{if}\qquad i=j \\
\max \frac{h^{k}_{ji}}{h^{k}_{ii}} \qquad
\textnormal{if} \qquad i\neq j. \end{array} \right.
 \end{eqnarray}
Since $\widehat{W}_{ij}^{\scriptsize{\textnormal{max}}}\leq \max
\varepsilon_{i}^{k} \bar{s}_{i}^{k}$, from (\ref{6.Proof of Proposition 2}) we have
\begin{eqnarray}\label{6.Proof of Theoream 1}
      \max_{\tilde{\mathbf{s}}_i\in \mathcal{R}_{s}} \rho (\textbf{W}^{\scriptsize{\textnormal{max}}})\leq  \|\overline{W}_{ij}^{\scriptsize{\textnormal{max}}} \|_{2}+ \|\overline{W}_{ij}(k)\|_{F}\leq  \|\overline{W}_{ij}^{\scriptsize{\textnormal{max}}}\|_{2}+ \sqrt{|M|}\|\mathbf{w}^{\scriptsize{\textnormal{max}}}\|_{2}.
\end{eqnarray}

\bibliographystyle{IEEEtran}
\bibliography{IEEEabrv,mybib}
\end{document}